\documentclass[pre,floatfix,twocolumn, 10pt]{revtex4} 
\usepackage{graphicx}
\usepackage{natbib}
\usepackage{dcolumn}
\usepackage{bm}
\usepackage{color}
\usepackage[colorlinks=true, linkcolor=blue, citecolor=blue]{hyperref}
\usepackage{subfigure}
\usepackage{graphicx,psfrag,xspace}
\usepackage{color}
\usepackage{amssymb,amsfonts,amsmath}
\usepackage{setspace}

\begin{document}

\author{E. A. Jagla} 
\affiliation{Centro At\'omico Bariloche, Instituto Balseiro, 
Comisi\'on Nacional de Energ\'ia At\'omica, CNEA, CONICET, UNCUYO,\\
Av.~E.~Bustillo 9500 (R8402AGP) San Carlos de Bariloche, R\'io Negro, Argentina}

\title{Down-hill creep of a granular material under expansion/contraction cycles}

\begin{abstract} 

We investigate the down-hill creep of a layer of granular material on a slope caused by an oscillatory variation of the size of the particles. The material is modeled as an athermal two dimensional polydisperse system of soft disks under the action of gravity. The slope angle is below the critical rest angle and therefore the system reaches an equilibrium configuration under static external conditions. However, under a protocol in which particles slowly change size in a quasistatic oscillatory way the system is observed to creep down in a synchronized way with the oscillation. We measure the creep advance per cycle as a function of the slope angle and the degree of change in particle size. 
In addition, we consider a situation in which the particle size oscillation amplitude decreases with depth, as it may be argued to occur in the case of a granular soil in an inclined terrain. In this case  creep profiles that are maximum at the surface and smoothly vanish with depth are obtained, as it is observed to occur in the field. 

\end{abstract}

\maketitle

\section{Introduction}

Soil slopes have a natural tendency to flow down hill in time, in order to reach more stable configurations. In some cases this manifests as abrupt rearrangements in the form of avalanches that occur during relatively short periods of time, involving the displacement of huge amounts of material across large distances.
Leaving aside these catastrophic events, there is a well documented tendency of  hill slopes to creep down as a function of time, in a slow process that becomes apparent only in typical periods in the scale of years.
\cite{young,kirkby}
 In these processes the upper layers of an inclined soil is observed to creep down, in such a way that vertically introduced tracer markers acquire a typical bent form, indicating that the surface layer creeps the most, while progressively deeper layers are less affected.\cite{roering} 

Models of hillslope creep is grounded in the work of Culling\cite{culling} who focused on the role of the porosity introduced by cycles of variation of ambient conditions (freeze–thaw, wet–dry, hot-cold). Eventually, the 
importance of cyclic processes in the creep behavior was well established, but the details of the process are not totally clear.
Beyond the continuum approach proposed by Culling and followed by others\cite{roering,kirkby,gabet,dietrich,roering2},
the importance to incorporate details of the
 grain-scale dynamics, has become progressively acknowledged.\cite{furbish,gray}

In recent years, theoretical and experimental advances on the understanding of granular materials\cite{nedderman,koenders} have opened new routes for the study of soil creep phenomena. Granular materials, as many kind of soils, can be described as yield stress fluids,\cite{coussot,berthier,ferrero-rmp} namely, materials that are solid and can withstand an applied stress if this is lower than some critical value, but that flow in a fluid-like way if this stress is overcome. The critical stress of these materials links very neatly with the rest angle in a hill (or heap) geometry, and typically it is expected that the observed angle in field be just slightly below the rest angle, in such a way that abrupt avalanche processes are nominally not able to occur. Yet, it is precisely in this sub-critical configuration that the soil creep process occurs.\cite{jerolmak}

One source of creep in yield-stress fluids is thermal activation. In fact, in the presence of stochastic thermal forces, energy barriers can be eventually surmounted, producing particle re-accommodations that would not be possible in the absence of thermal effects.\cite{13,14,15} This generates that the material is able to weakly flow even if below its critical stress.\cite{js,fkj} The flow rate caused by thermal activation is typically much smaller than that in the true flow regime (above $\sigma_c$), and this is one of the characteristics of what is called a creep regime. The effect of thermal fluctuations is intimately related to the size of the elemental constituents in the system. It can have a sizeable effect
in cases in which the elementary grain size is very small, as for instance in metallic glasses, but its effect becomes less and less important as the grain size increases. In granular materials with grain sizes larger than about tens of microns it is hard to justify that it plays any role. 
Therefore, it is generally accepted that thermal activation is not a crucial ingredient in the dynamics of soil slopes. 

Mechanical perturbations are natural candidates to be considered as a possible origin of the creep phenomenon. For instance, vibrations caused by a variety of sources such as walking of animals, movement of vegetation caused by wind, water falling and flow during rain, and even earthquakes, all have the potential to produce creep to some extent.\cite{bontemps,auzet,fleming,eyles,matsuoka} These sources of mechanical perturbations share some similarities with thermal fluctuations, namely that they all have some degree of stochasticity. While this may be an important source of creep, it is not the one on which we will focus in the present work.

The kind of perturbation giving rise to creep that will be considered here
is the variation of one global parameter in the system.
The idea of subcritical creep induced by cyclic variations of one parameter was recently presented in a mesoscopic model of a yield-stress material in \cite{ferrero_jagla_unp}. 
It was shown in there that the periodic variation of the stiffness of an elastic surface that lays on top of an inclined, disorder energy potential can induce creep, even if the surface would stay completely at rest in the absence of such a perturbation.
Here we apply this idea to a
model granular material in which the size of the particles changes slowly in time in an oscillatory way, mimicking an expansion/contraction behavior caused for instance bay a day/night variation of humidity, or temperature\cite{temperature}. 
We implement this idea in a model two-dimensional granular material formed by a poly-disperse collection of discs, through molecular dynamics simulation.
We set subcritical conditions, namely in a configuration in which the system reaches a long lasting stable configuration if parameters are kept constant.
However, if the size of the particles is changed periodically in time, 
a creep behavior correlated with the cyclic variation of size is observed.
We provide quantitative details of this process, in particular how the creep rate depends on the oscillation amplitude of particle size, and on the closeness of the material free slope to the rest angle.

\section{Numerical details}

We model a granular material under the action of gravity, with a free surface forming with the horizontal an angle $\theta$ lower than the critical rest angle $\theta_0$. 
The numerical model we use is a two-dimensional system of polydisperse circular particles interaction through two body central forces. 
Radius of the particles are chosen from a uniform distribution in the range between $r_{min}$ and $r_{max}=2r_{min}$. The interaction potential between particles $i$, $j$ (with radius $R_i$, $R_j$) is described by a purely repulsive potential  of the form 
\begin{eqnarray}
V_{ij}=V_0 [d-(R_i+R_j)]^2~~~~~~~\mbox {if}~~~~~~  d<(R_i+R_j)\nonumber\\
V_{ij}=0       ~~~~~~~\mbox {if}~~~~~~  d>(R_i+R_j)
\label{pot}
\end{eqnarray}
where $d$ is the separation between particles, and $V_0$ sets the energy scale.
From now on, we use dimensionless units setting $r_{min}=1$, $V_0=1$.
The time evolution of the system is a fully overdamped dynamics where the instantaneous velocity of each particle is proportional to the total force acting on it:
\begin{equation}
\frac{d{\bf r}_i}{dt}=-\sum_{j\ne i}\frac{dV_{ij}}{d{\bf r}_j}+ \pi R_i^2{\bf g}
\label{g}
\end{equation}
The force acting on the particles includes the one originated in the interparticle interaction potential, and that coming from a  gravity acceleration $\bf{g}$, considering particles of the same density
to avoid differential buoyancy effects. 
This relaxational dynamics is appropriate in view of the quasi-static conditions of the problem.

As the system is supposed to be located on an inclined plane with an angle below the rest angle (see the implementation below), it eventually reaches a fully stable equilibrium configuration. On this configuration quasistatic variation of the particle radius are applied as explained below, and this drives the slow creep of the system.  
This basic model is applied in two different situations, that we call the {\em homogeneous} simulation, and the {\em  layered} simulation configuration.

\begin{figure}
\includegraphics[width=5cm,clip=true]{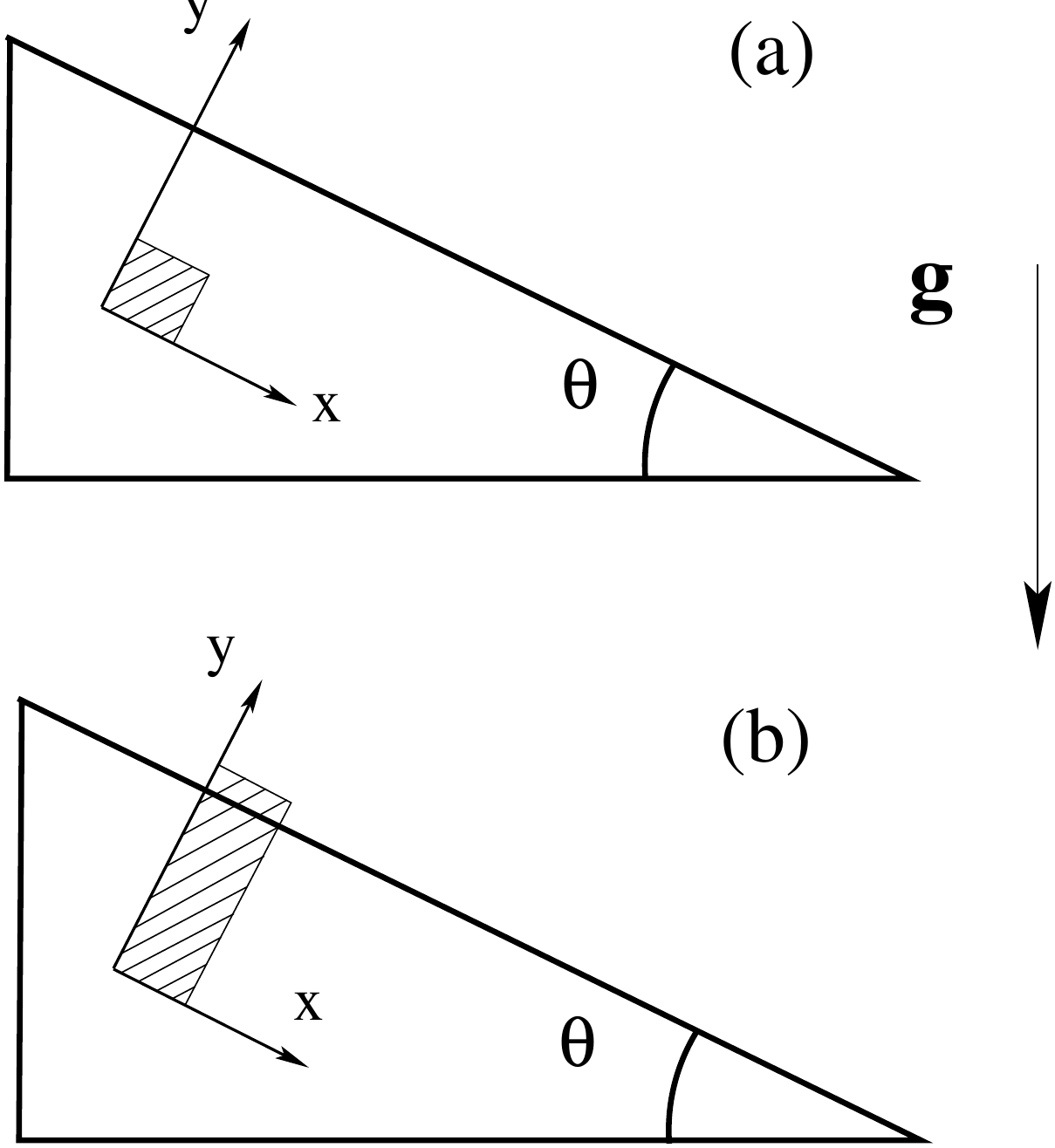}
\caption{The triangle represents a granular system with the surface having a finite slope $\theta$. Dashed regions are the pieces actually simulated in the homogeneous (a) and layered (b) simulation protocols.}
\label{f1}
\end{figure}

\section {Results}
\subsection{Homogeneous simulations}

In the homogeneous configuration we intend to model a small homogeneous portion of material at a fixed depth  (Fig. \ref{f1}(a)).
We do so by orienting the simulation box  (of size $l_x\times l_y$) with the $x$ axis along the downslope direction, and $y$ axis perpendicular to the free surface. As the simulation box represents a small piece of a homogeneous material we use in this case periodic boundary conditions and disregard the 
$y$ component of gravity, since it is a homogeneous force. The $x$ component of gravity represents a shear stress  in the system that we include in the form of Lees-Edwards boundary condition along $y$. To this end, calling $\delta$ the shift in the boundary condition, we use the following scheme: the actual stress value $\sigma_m$ measured along the simulation is used to update the shift $\delta$ according to 
\begin{equation}
\frac{d\delta} {dt}=\eta (\sigma-\sigma_m)
\end{equation}
where $\sigma$ is the target stress value we want to reach, and $\eta$ is an appropriately chosen convergence factor ($\eta=0.03$ in the simulations below). 
The shift $\delta$ determines the strain $\varepsilon$ in the system as $\varepsilon=\delta/l_y$. 
As $\sigma$ is supposed to be lower than the critical value $\sigma_c$, the system is guaranteed to reach a stable configuration (with a constant $\varepsilon$) and would stay there forever in the absence of additional external perturbations. 

 Such a perturbation is introduced in the form of an oscillatory (homogeneous) change in the size of the particles, that pass from original values $R_i$ 
to maximum final values $R_i(1+\Delta)$ (with $\Delta>0$), and then return to their original sizes, the process being repeated in time. In the homogeneous geometry of the simulation (Fig. \ref{f1}(a)), an increase in particle size implies an increase of normal stress along the $x$ direction, that induces an expansion along $y$ (allowed by the existence of the free surface). The process can be modeled by effectively keeping the particle size as fixed, but reducing the length $l_x$ down in a factor $1+\Delta$, and increasing the $l_y$ size by the same factor so as to keep the average system density as fixed. This is the numerical protocol that is implemented. 
Under this process, and keeping the applied stress value $\sigma$ as fixed, the value of the strain $\varepsilon$ increases with the cyclic variation of $\Delta$. The average increase  of $\varepsilon$ on each cycle, notated $\delta\varepsilon$, is the main outcome of the simulations. 

\begin{figure}
\includegraphics[width=9cm,clip=true]{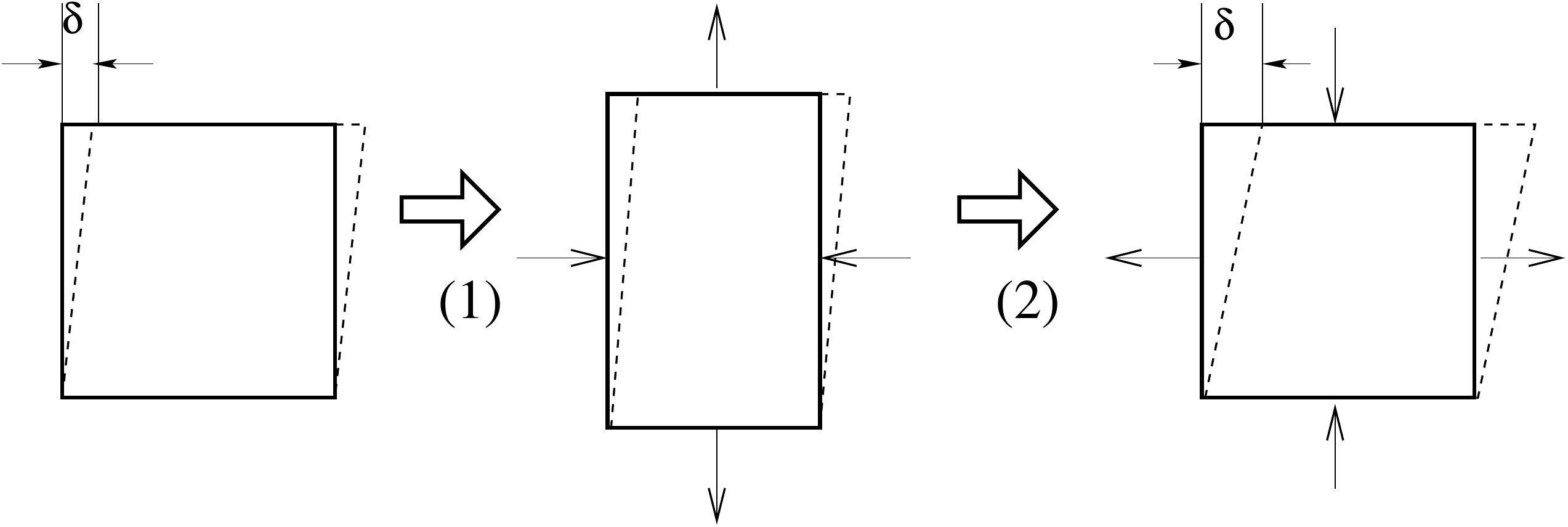}
\caption{One cycle of the process of increasing particle size (step 1, effectively modeled by reshaping the simulation box, see text for explanation), and decreasing back to the original situation (step 2). In the process, the value of shift $\delta$ in boundary conditions needs typically to be increased in order to maintain the same value of $\sigma$.}
\label{f2}
\end{figure}

\begin{figure}
\includegraphics[width=9cm,clip=true]{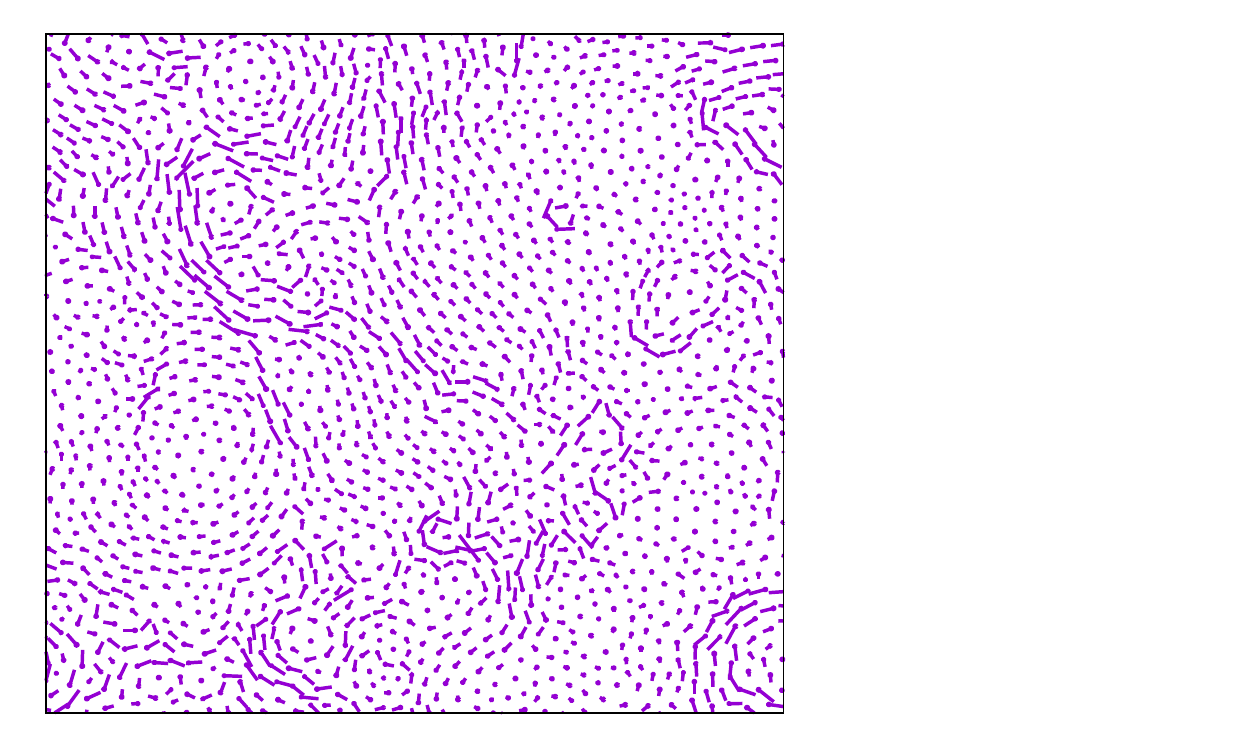}
\caption{Non-affine displacements of the particles during a full cycle of the homogeneous simulation, at $\sigma=.0045$, $\Delta=.06$.}
\label{snap}
\end{figure}

We simulate a system of 2000
particles, that are initially placed randomly in a box of size $l_x=l_y=125$.
The configuration is relaxed until all particles reach a stable position.
We first apply an equilibration protocol by applying a stress $\sigma$ such that the system yields
permanently (i.e., $\delta$ does not reach a constant equilibrium value, but increases linearly in time). Then $\sigma$ is reduced in small steps
until we detect that $\delta$ remains constant, meaning that we have just reached (from above) the critical stress value $\sigma_c$. It turns out that $\sigma_c\simeq 0.009$ in our case. \cite{foot1}
Then a value of stress $\sigma<\sigma_c$ is applied, and the oscillation protocol is started:
the simulation box is deformed from its original square form ($l_x=l_y=125$) to
a rectangular shape
[$l_x(1+\Delta)$, $l_y/(1+\Delta)$, with typical values of $\Delta$ up to 0.15] at a slow constant rate, keeping the applied shear stress $\sigma$ as constant (Fig. \ref{f2}, step 1). 
Once the final configuration is reached, and no particle reacommodations are detected,  the box is returned back to its original shape (Fig. \ref{f2}, step 2). In this cyclic process, plastic particle rearrangements can occur as Fig. \ref{snap} shows, and this produces that the applied shift $\delta$ (and strain $\varepsilon=\delta/l_y$) typically increases after each cycle in order to maintain the shear stress constant.
 
\begin{figure}
\includegraphics[width=7cm,clip=true]{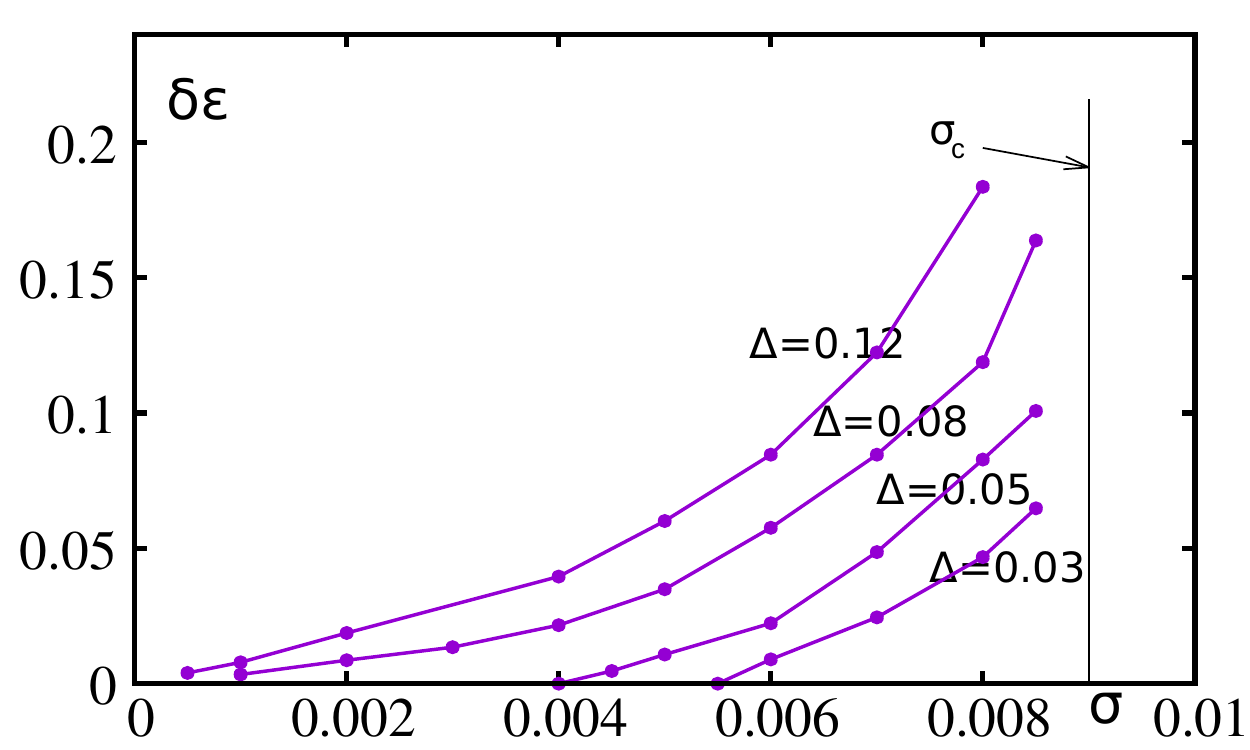}
\caption{Average increase in strain per cycle $\delta \varepsilon$ as a function of the constant applied stress $\sigma$, for different values of the deformation parameter $\Delta$, in the case of homogeneous simulations.}
\label{f4}
\end{figure}

The value of strain $\varepsilon$ is recorded during many cycles of the process.
Results obtained for the average strain increase per cycle (or creep rate) $\delta \varepsilon$ for different values of $\sigma$ and $\Delta$ can be seen in Fig. \ref{f4}. We observe in this figure systematic trends that coincide with the results found in \cite{ferrero_jagla_unp} using an effective mesoscopic model of cyclic perturbations. The main points are the following. At a fixed value of the oscillation amplitude $\Delta$, creep rate is an increasing function of $\sigma$, becoming very large as $\sigma\to\sigma_c$ (where the system tends to flow even in the absence of oscillation), and reducing to zero at some lower stress $\sigma_0<\sigma_c$. The value of $\sigma_0$ vanishes if  $\Delta$ is large enough ($\Delta\gtrsim 0.1$ in our case). \cite{breathing} On the other hand, if $\Delta$ is too small ($\Delta\lesssim 0.02$)  we have observed in the simulations that although creep occurs during a few cycles, eventually the system reaches a perfectly periodic behavior and $\delta\varepsilon$ vanishes. Based on the results of \cite{ferrero_jagla_unp}, we believe this is related to the finite size of the system, and expect that for larger system sizes some creep exists for any non-zero $\Delta$, although only very close to $\sigma_c$. 


Simulations in the homogeneous configuration give a thorough description of how the material responds locally, under a given degree of variation in particle size. 
However, it is important also to consider the combined effect of different layers of material experiencing different degree of particle size variations. This is the aim of the layered simulations presented in the next Section.

\subsection{Layered simulations}

In the layered configuration (Fig. \ref{f1}(b)) we intend to model a thick piece of material, from surface to a deep layer that is supposed not to be affected by the oscillatory change in ambient conditions that drive the variation of particle size.
The $x$ boundary conditions continue to be periodic, but
now the conditions along $y$ are free, and the $y$ component of gravity is explicitly included. This means that a free surface naturally appears. At $y=0$ in the simulation box (the bottom layer) a rigid wall that particles cannot penetrate is assumed. In addition, a buffer layer of particles between $y=0$ and $y=y_b=10$ is included, in such a way that these particles remain fixed, and serve as a connection of the upper part of the system to the frozen bulk.
The $x$ component of gravity is explicitly included as a constant $x$ force on each particle. Notice however that particles with $y<y_b$ do not move at all, therefore avoiding a uniform slip of the whole system. The parameter $\Delta$ quantifying the change in particle size is now $y$-dependent, namely we have a $\Delta(y)$ function. The justification of such a dependence is that it is expected that external conditions driving the change of particle size act from above the system, and their effect is progressively reduced in deeper layers. Therefore we will use for the $\Delta(y)$ function  a monotonous function that is maximum at the surface and vanishes at some finite depth.
For concreteness we use a $\Delta(y)$ function that decays linearly from the surface (where it attains its maximum value $\Delta_0$) to 0 at some depth $y_0>y_b$. In this way  we can evaluate to what extent the system responds locally to the degree of change in particle size, or if there are important effects that can be ascribed to a non-local rheology (as would be the case for instance if creep below $y_0$ is observed).

We use a system with $N=2000$ particles, in a simulation box with $l_x=125$. To obtain the starting sample we drop the particles randomly and allow them to settle under the vertical component of gravity force. After this initial stage, the positions of particles below $y_0=10$ are set as fixed, and full
gravity force is applied. ${\bf g}$ (Eq. \ref{g}) has now $x$ and $y$ components, such as $g_x/g_y=\tan\theta$, where $\theta$ is the slope angle of the surface.
We use a fixed value of $g_y$, namely $g_y=0.01$.
Applying progressively larger values of $g_x$, and
after some local reaccommodations, the system reaches stable configurations  up to values of $g_x/g_y\lesssim 0.09$. This corresponds to a "rest angle" $\theta_0\simeq 5 ^\circ$. \cite{foot2}.
We therefore restrict to values of $g_x \lesssim 0.09 g_y=0.0009$ to define the starting sub-critical system.

\begin{figure}
\includegraphics[width=2.3cm,clip=true]{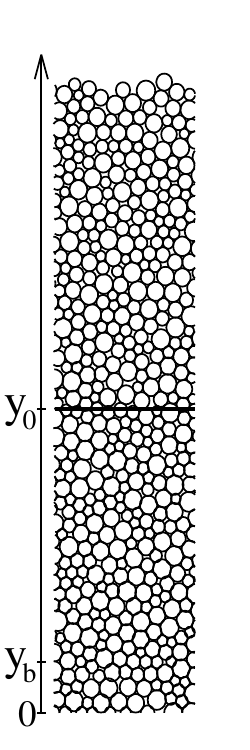}
\includegraphics[width=2.3cm,clip=true]{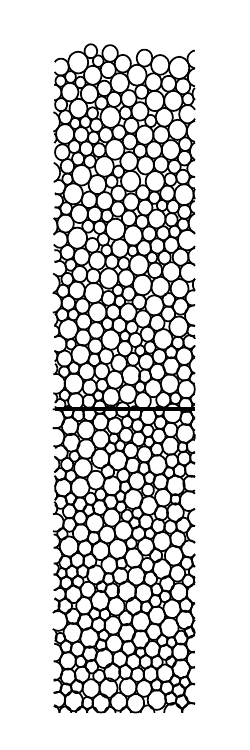}
\includegraphics[width=2.3cm,clip=true]{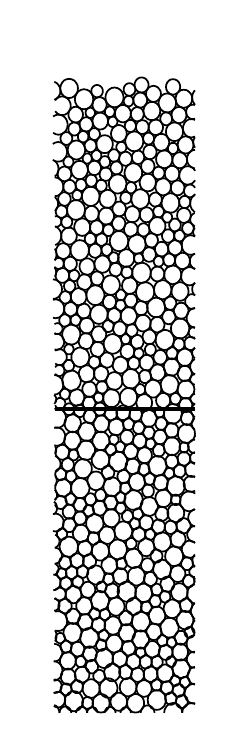}
\includegraphics[width=7cm,clip=true]{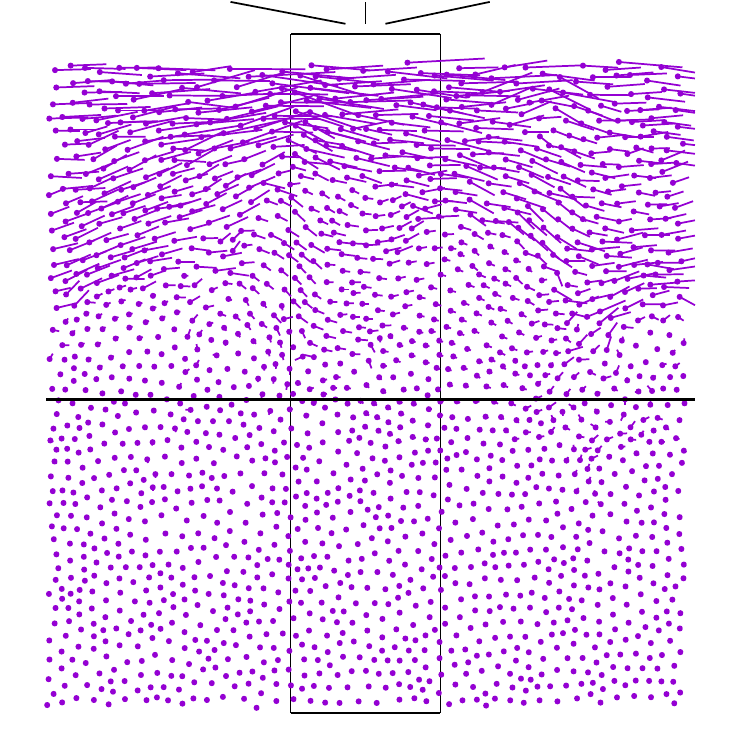}
\caption{
(a) The actual particles within the rectangular box in (b) at the beginning of the cycle, at maximum expansion, and after contraction.
(b) Particles at the beginning of the cycle in the full simulation box (dots) and their displacements 
in one expansion/contraction cycle (segments).}
\label{breath}
\end{figure}

Figure \ref{breath}(a) displays 
the actual particles in a portion of the system (the rectangular box in panel (b)) at the beginning of the cycle, 
after expansion, and after final contraction.
The vertical down-force is $g_y=0.01$, and the horizontal force is $g_x=0.0008$. The degree of expansion/contraction  $\Delta$ is zero below $y_0$ and increases linearly up to $\Delta=0.1$ at the surface. Also the layer between $y=0$ and $y_b$ corresponds to the buffer layer, where particles are not allowed to move at all.
The swelling of the system caused by the increase in particle size is clearly seen in the uplift observed in the configuration after expansion. 
Panel (b) shows
the change in particle positions
after the full expansion/contraction cycle. 
A systematic shift in the $+x$ direction is clearly visible. The effect is the largest at the surface, and decreases in depth. Note that although particles below the black line were not affected by size change, some particles below this line suffered a change in their horizontal position influenced by the position change of particles above them.  

\begin{figure}
\includegraphics[width=7cm,clip=true]{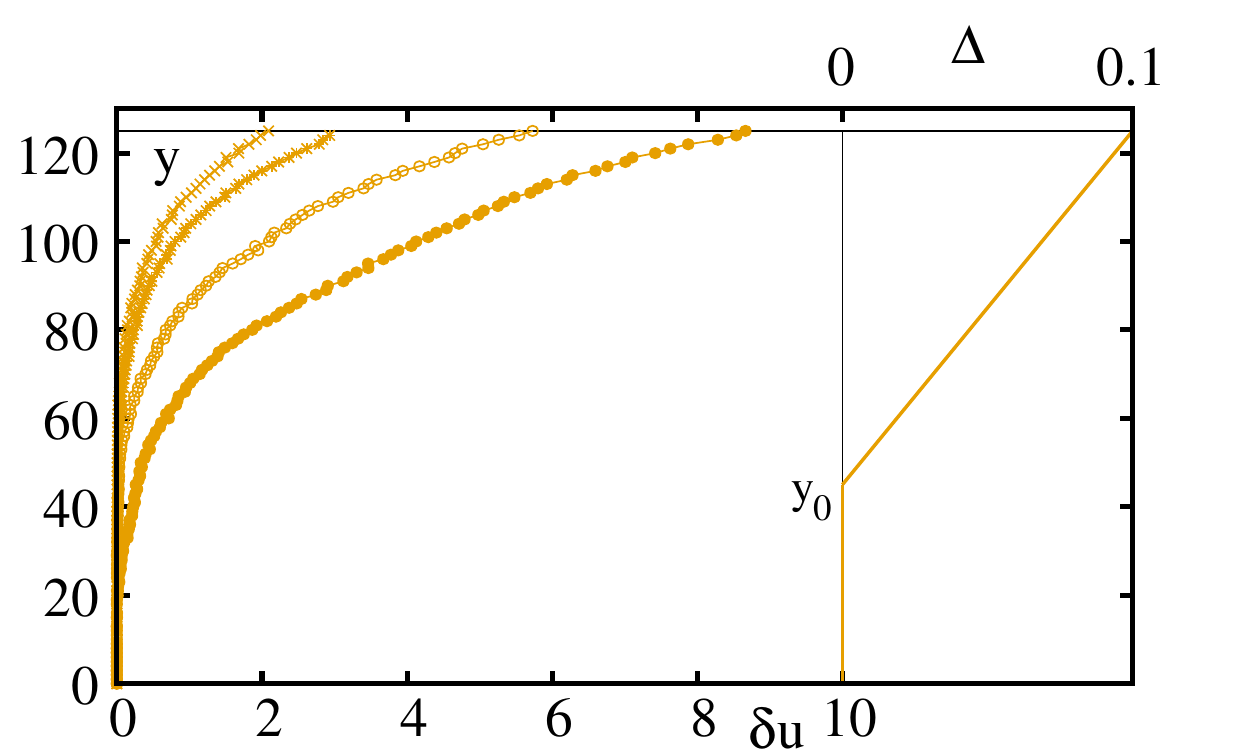}
\includegraphics[width=7cm,clip=true]{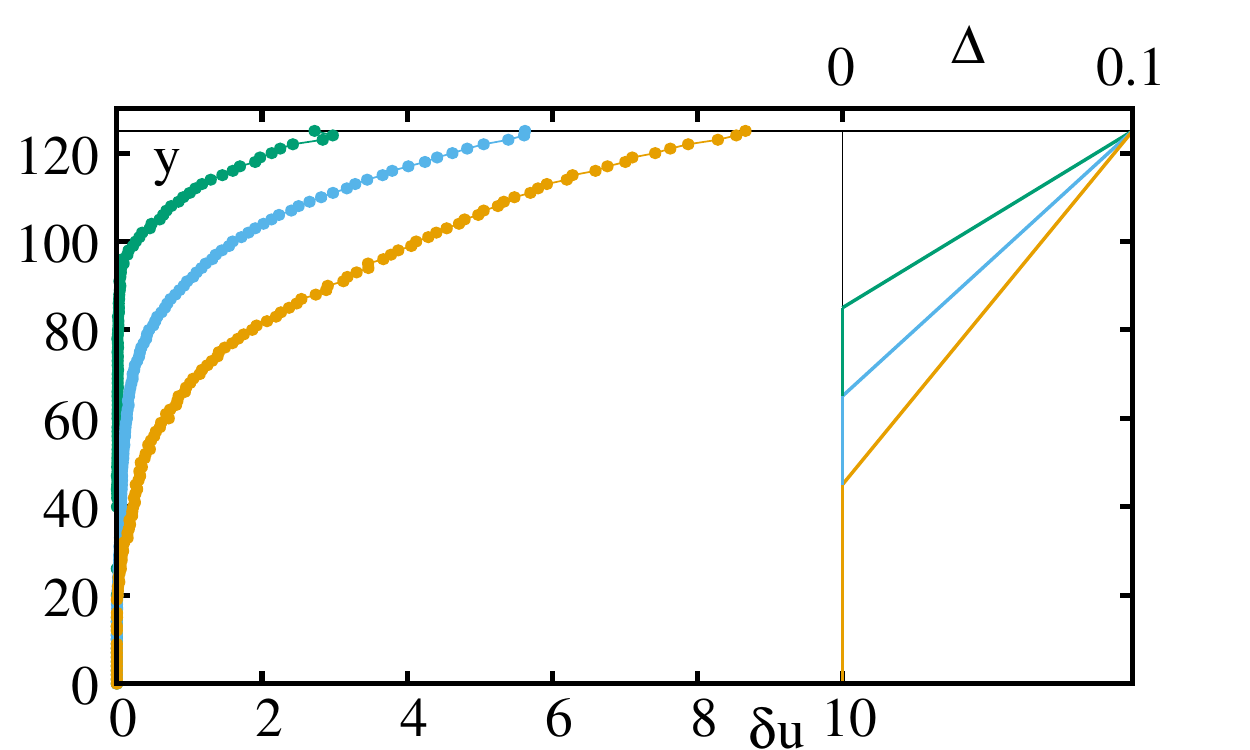}
\caption{(a) Advance rate $\delta u$ in units of particle radius per cycle (averages over 100 cycles) as a function of depth in the system  when particle change their radios in the factor $1+\Delta$, with $\Delta(y)$ as indicated in the right part of the plot. Different curves correspond to different horizontal forces ($g_x=0.0008$, 0.0007, 0.0006, 0.0005, from right to left curves). (b) Same as panel (a) but keeping $g_x=0,0008$ as fixed, and varying the width of the surface layer on which breathing is applied.}
\label{expo}
\end{figure}

Running this expansion/contraction process during many cycles for different values of $\Delta_0$ and $y_0$, we constructed Fig. \ref{expo}. The points represent the profile of creep advance $\delta u$ (i.e., the increase in the $x$ coordinate of the particles per cycle) as a function of depth. 
In panel (a) we show four curves, obtained by averaging over 100 cycles of increase/decrease particle size, for a value of $\Delta_0=0.1$, $y_0=45$, and different values of $g_x$, corresponding to different values of the slope angle $\theta$. We observe a creep profile that is maximum at the surface and progressively decreases in depth. However, there is no strict vanishing of the creep at the depth at which the particle size does not oscillate. An attempt to obtain creep at a value of $g_x=0.0004$ eventually produced a cyclically stable configuration, namely, a zero advance per cycle. In panel (b) we keep the same value of $g_x=0.0008$ in the three curves, but vary the value of $y_0$ above which breathing is applied. The profile of $\Delta(y)$
is indicated in the rightest part of the figure. Also in this case, if $y_0$ is increased above $\sim 90$ the cyclic advance vanishes.

\begin{figure}
\includegraphics[width=7cm,clip=true]{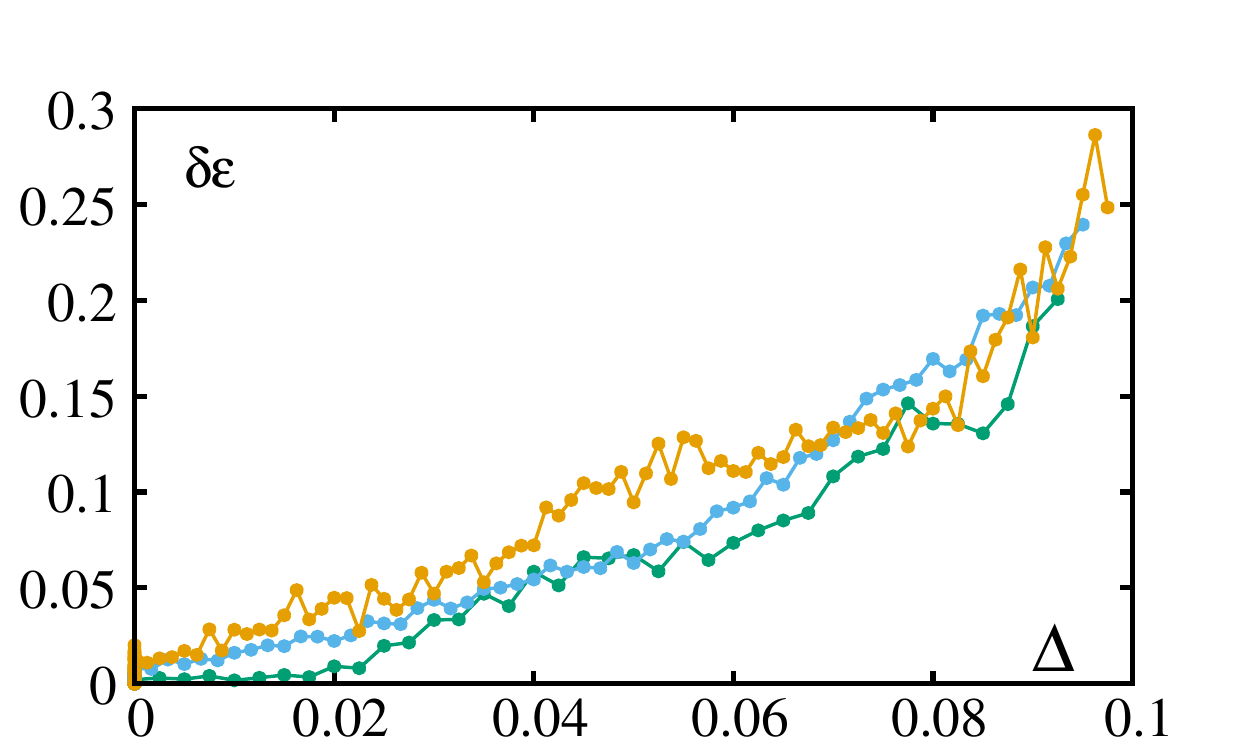}
\caption{Results for  $\delta \varepsilon\equiv d(\delta u)/dy$, calculated for the three curves in Fig. \ref{expo}(b), and plotted against the local value of $\Delta$. The approximate coincidence of the three curves indicated that local rheology behavior is fulfilled to a good extent.}
\label{deriv}
\end{figure}

The profiles in Fig. \ref{expo} have a strong resemblance of experimental results in different kind of soils. \cite{young,kirkby,roering}
Of course, it can be argued that we are setting here by hand a linear variation profile $\Delta(y)$, and it may be argued that other forms of this function could produce different forms of $\delta u(y)$. Yet we notice that the profiles in
Fig. \ref{expo} are to a good extent compatible with a local rheology behavior in which the local strain creep rate $\delta \varepsilon\equiv d(\delta u)/d y$ responds to the local value of $\Delta$ (and the applied value of $g_x$). For instance, the data in Fig. \ref{expo}(b) can be replotted by numerically calculating $d(\delta u)/d y$ and plotting it against the local $\Delta$ value (notice that $g_x$ is the same in all three curves). The result, shown in Fig. \ref{deriv} shows a rather good collapse, indicating that the local rheology assumption is rather good. We notice that a small degree of non-local behavior is observed in the cases in which we observed creep in layers that do not expand. This residual creep is induced by layers expanding some distance above. 
Local rheological behavior thus implies that an observed profile of $\varepsilon(y)$ should correlate with the local degree of expansion through 
\begin{equation}
\delta \varepsilon \equiv \frac{d(\delta u)}{dy} =F (\Delta(y))
\end{equation}
where the form of the $F$ function can be read out from the homogeneous results in Fig. \ref{f4}.
The usual quasi-exponential profiles observed in field can thus be considering as appearing 
from a combination of the form of the function $F(\Delta)$, and the perturbation profile $\Delta(y)$.
In any case, it is interesting to note that in the present model the typical width of the surface layer affected by creep corresponds to the width of the surface layer that experiences appreciable expansion/contraction effect. This fact seems to be amenable to experimental verification.

\section{Summary and conclusions}

In this paper we have presented numerical simulations of a two-dimensional system of poly-disperse soft disks, interacting through two-body central forces, under the action of gravity. The aim was to model the behavior of 
a heap of material or a hill slope of terrain and the possibility that it experiences slow creep motion driven by a periodic variation of ambient conditions. Change of external conditions was effectively simulated by 
changing the size of the particles quasistatically in time, in a periodic way. Since we work at slope angles $\theta$ smaller than the rest angle $\theta_0$ of the model, at any step in the simulation process the system can achieve a static equilibrium configuration (if particle sizes are not changed). However, our main result is that after a complete expansion contraction cycle the system experiences some degree of creep in the downhill direction. 

Quantitative results have been provided for two different set ups.
In the first case of homogeneous simulations we focus on a small portion of soil, at a given depth, under the action of a finite shear stress, yet below the critical value necessary to enter the flow regime. On this configuration a periodic and homogeneous increase/decrease in size of the particles is applied and we observe how the system yields a finite amount on each cycle, therefore justifying the existence of a sub-critical creep caused by the oscillation of particle sizes. The creep rate is investigated in this case as a function of the degree of change in particle size, and the value of stress.

In the second set of results, we model a full thick layer of soil, from free surface to an unperturbed layer in depth, under the action of the same   increase/decrease size of the particles now acting only on a superficial layer, i.e., more akin to a realistic situation where humidity or temperature variations act from the free surface, and its effect progressively decays in depth. The main results in this case indicate that the system responds mostly locally, i.e., the strain increase per cycle at any given depth is a function of the local degree of particle size change, the same that was obtained in the homogeneous simulations. The main outcome of these results is that they clarify the fact that the thickness of the surface layer that yields corresponds to the layer that is affected by the perturbation. In other words, non-local effects in which a perturbed surface layer produces yielding in layers well below it do not occur in our simulations. Non-local effects are limited to small distances of a few particle radius.

\section{Acknowledgments}

I thank E.E. Ferrero for stimulating discussions on this topic.


\begin{thebibliography}{1}
\bibitem{temperature} The thermal expansion we mention here has obviously nothing to do with temperature producing stochastic forces onto the particles. Stochastic forces associated with temperature are totally neglected in our approach.
\bibitem{foot1}This stress value depends strongly (and almost linearly) on the hydrostatic pressure $P$ in the system, which itself is highly dependent on density. In the present simulations the particle density is 0.128, and the pressure turns out to be $P\simeq 0.11$, therefore $\sigma_c/P\simeq 0.082$ that correlates very well with the ratio $g_x/g_y$ determining the rest angle in the layered simulations of Section IIIB.
\bibitem{breathing}The case in which $\sigma_0=0$ can be described as a "fluidized" case, in which the system creeps even for a very small applied $\sigma$. This can be connected with the analysis in \cite{berthier2} where a oscillation in particle sizes (but for the case in which phases of the oscillation are not the same in all particles) was analyzed as a model for active matter.
\bibitem{foot2} This value looks very small compared with realistic values of $\theta_0$ of sand heaps, for instance. The main reason of this small angle is the use of particles without friction. 
This could in principle be cured by the introduction of friction, or some kind of roughness of the particles.
\end{thebibliography}
\end{document}